\documentclass[english]{revtex4}
\usepackage[T1]{fontenc}
\usepackage[latin9]{inputenc}
\usepackage{esint}
\usepackage{babel}

\begin{document}

\title{Thermodynamics of the Three-dimensional Black Hole with a Coulomb-like
Field}

\author{Alexis Larrañaga}

\affiliation{Universidad Nacional de Colombia. Observatorio Astronómico Nacional
(OAN)}

\affiliation{Universidad Distrital Francisco José de Caldas. Facultad de Ingeniería.
Proyecto Curricular de Ingeniería Electrónica}

\email{ealarranaga@unal.edu.co}

\author{Luz Angela García}

\affiliation{Universidad Nacional de Colombia. Departamento de Física }

\email{lagarciape@unal.edu.co}

\begin{abstract}
In this paper, we study the thermodynamical properties of the $\left(2+1\right)$
dimensional black hole with a Coulomb-like electric field and the
differential form of the first law of thermodynamics is derived considering
a virtual displacement of its event horizon. This approach shows that
it is possible to give a thermodynamical interpretation to the field
equations near the horizon. The $\Lambda=0$ solution is studied and
its interesting thermodynamical properties are commented.
\end{abstract}
\maketitle
As is well known, the electric field of the BTZ black hole is proportional
to the inverse of $r$, hence its potential is logarithmic. If we
are interested in a solution with a Coulomb-like electric field (proportional
to the inverse of $r^{2}$), we need to consider non-linear electrodynamics.
This kind of solution was reported by Cataldo et. al. \cite{cataldo},
and describes charged-AdS space when considering a negative cosmological
constant. 

The thermodynamical properties of black holes are associated with
the presence of the event horizon. In particular, Jacobson\cite{jacobson}
and Padmanabhan \cite{padman} established that the first law in
differential form,

\begin{equation}
dM=TdS+\Omega dJ+\Phi dQ,\end{equation}
 can be obtained from the Einstein's field equations by using the
idea of a virtual displacement of the horizon. The same idea was applied
by Akbar \cite{akbar} and Akbar and Siddiqui \cite{akbar2} to
the BTZ black holes to show that the thermodynamical interpretation
of the field equations holds for the static and non-static BTZ metrics
in $\left(2+1\right)$ gravity. 

In this paper we investigate the thermodynamics of the three-dimensional
black hole with a nonlinear electric field reported in \cite{cataldo},
to show that the field equations include the first law of thermodynamics
in differential form. We also consider the $\Lambda=0$ black hole
to show that it has interesting thermodynamical properties.

\section{The Black Hole Solution}

The metric reported by Cataldo et. al. \cite{cataldo} is a solution
of the $\left(2+1\right)$ dimensional Einstein's field equations
with a negative cosmological constant $\Lambda=-\frac{1}{l^{2}}<0$,

\begin{equation}
G_{\mu\nu}-\frac{g_{\mu\nu}}{l^{2}}=\pi T_{\mu\nu},\end{equation}
where we have used units such that $G=\frac{1}{8}$. As is well known,
the electric field for a static circularly symmetric solution in three
dimensions (charged BTZ solution \cite{btz}) is proportional to
the inverse of $r$, i.e.

\begin{equation}
E\propto\frac{1}{r},\end{equation}
and therefore, the potential is logarithmic,

\begin{equation}
A\propto\ln r.\end{equation}

To obtain a different electric field, Cataldo et. al. used a nonlinear
electodynamics. In the non-linear theory, the action $I$ does not
depend only on the invariant $F=\frac{1}{4}F_{\mu\nu}F^{\mu\nu}$,
but it can be a generalization of it, for example

\begin{equation}
I\propto\int d^{3}x\sqrt{\left|g\right|}\left(F_{\mu\nu}F^{\mu\nu}\right)^{p},\end{equation}
where $p$ is some constant exponent. If the energy-momentum tensor
is restricted to be traceless, the action becomes a function of $F^{3/4}$,
and the static circularly symmetric solution obtained has the line
element 

\begin{equation}
ds^{2}=-f\left(r\right)dt^{2}+\frac{dr^{2}}{f\left(r\right)}+r^{2}d\varphi^{2},\label{eq:staticmetric}\end{equation}
where

\begin{equation}
f\left(r\right)=-M+\frac{r^{2}}{l^{2}}+\frac{Q^{2}}{6r}.\end{equation}

The electric field for this solution is

\begin{equation}
E\left(r\right)=\frac{Q}{r^{2}},\label{eq:electric field}\end{equation}

which is the standard Coulomb field for a point charge. The metric
depends on two parameters $Q$ and $M$, that are identified as the
electric charge and the mass, respectively.

\subsection{Horizons}

The horizons of this solution are defiened by the condition \begin{equation}
f\left(r\right)=0\end{equation}
 or

\begin{equation}
-M+\frac{r^{2}}{l^{2}}+\frac{Q^{2}}{6r}=0,\label{eq:horizonmass}\end{equation}
that can be transformed into a third-order polynomial,

\begin{equation}
r^{3}-\left(Ml^{2}\right)r+\frac{Q^{2}l^{2}}{6}=0.\label{eq:3orderpol}\end{equation}

This polynomial (\ref{eq:3orderpol}) can be written as

\begin{equation}
r^{3}+pr+q=0,\end{equation}
where

\begin{eqnarray}
p & = & -Ml^{2}=\frac{M}{\Lambda}\\
q & = & \frac{Q^{2}l^{2}}{6}=-\frac{Q^{2}}{6\Lambda}.\end{eqnarray}
To obtain the roots of this polynomial we need to establish a classification
criteria based on the parameter

\begin{equation}
H=\left(\frac{p}{3}\right)^{3}+\left(\frac{q}{2}\right)^{2},\end{equation}

and using the quantity

\begin{equation}
R=\mbox{sign}\left(q\right)\sqrt{\frac{\left|p\right|}{3}}=\sqrt{\frac{Ml^{2}}{3}}.\end{equation}
With these definitions the roots are parameterized by the auxiliarly
angle $\phi$, that depends on the values of $p$ and $H$, and we
can establish three cases.

\subsubsection{Case I. $p<0$ and $H\leq0$.}

In this case we have a negative cosmological constant, $\Lambda<0$
and 

\begin{equation}
\left(\frac{-Ml^{2}}{3}\right)^{3}+\left(\frac{Q^{2}l^{2}}{12}\right)^{2}\leq0\end{equation}

\begin{equation}
\frac{Q^{4}}{16}\leq\frac{M^{3}l^{2}}{3}.\end{equation}
Therefore, the condition for this case is given, in terms of the cosmological
constant, as

\begin{equation}
-\frac{16M^{3}}{3Q^{4}}\leq\Lambda<0.\end{equation}
Under this condition, the auxiliary angle is defined by

\begin{equation}
\cos\phi=\frac{q}{2R^{3}}\end{equation}

or

\begin{equation}
\cos\phi=\frac{Q^{2}}{4Ml}\sqrt{\frac{3}{M}}\end{equation}

and the roots are all real,

\begin{eqnarray}
r_{1} & = & -2R\cos\left(\frac{\phi}{3}\right)\\
r_{2} & = & -2R\cos\left(\frac{\phi}{3}+\frac{2\pi}{3}\right)\\
r_{3} & = & -2R\cos\left(\frac{\phi}{3}+\frac{4\pi}{3}\right).\end{eqnarray}

\subsubsection{Case II. $p<0$ and $H>0$.}

This time we also have a negative cosmological constant, $\Lambda<0$,
but

\begin{equation}
\left(\frac{-Ml^{2}}{3}\right)^{3}+\left(\frac{Q^{2}l^{2}}{12}\right)^{2}>0,\end{equation}
which means

\begin{equation}
\frac{Q^{4}}{16}>\frac{M^{3}l^{2}}{3}.\end{equation}
Then, the cosmological constant must be such that

\begin{equation}
\Lambda<-\frac{16M^{3}}{3Q^{4}}.\end{equation}
In this case, the auxiliary angle is defined by

\begin{equation}
\cosh\phi=\frac{q}{2R^{3}}\end{equation}

or

\begin{equation}
\cosh\phi=\frac{Q^{2}}{4Ml}\sqrt{\frac{3}{M}}\end{equation}

and the roots are

\begin{eqnarray}
r_{1} & = & -2R\cosh\frac{\phi}{3}\\
r_{2} & = & R\cosh\frac{\phi}{3}+i\sqrt{3}R\sinh\frac{\phi}{3}\\
r_{3}=r_{2}^{*} & = & R\cosh\frac{\phi}{3}-i\sqrt{3}R\sinh\frac{\phi}{3}.\end{eqnarray}

\subsubsection{Case III. $p>0$ and $H>0$.}

Since $M>0$, in this case the cosmological constant must be positive,
$\Lambda>0$, and

\begin{equation}
\left(\frac{M}{3\Lambda}\right)^{3}+\left(\frac{Q^{2}}{12\Lambda}\right)^{2}>0\end{equation}

or

\begin{equation}
\Lambda>0>-\frac{16M^{3}}{3Q^{4}}.\end{equation}
Since the cosmological constant in this case is positive, this condition
is already fulfilled. The auxiliary angle is defined by

\begin{equation}
\sinh\phi=\frac{q}{2R^{3}}\end{equation}

or

\begin{equation}
\sinh\phi=\frac{Q^{2}}{4M}\sqrt{\frac{3\Lambda}{M}}\end{equation}

and the roots are

\begin{eqnarray}
r_{1} & = & -2R\sinh\frac{\phi}{3}\\
r_{2} & = & R\sinh\frac{\phi}{3}+i\sqrt{3}R\cosh\frac{\phi}{3}\\
r_{3}=r_{2}^{*} & = & R\sinh\frac{\phi}{3}-i\sqrt{3}R\cosh\frac{\phi}{3}.\end{eqnarray}
\\
\\
For the black hole solution studied in this paper, the relevant cases
are I and II, because the cosmological constant must be negative (i.e.
we consider the charged-AdS space). In case I we have three real horizons,
while in case II we have just one real horizon. 

Note that in case I the limit condition $H=0$ defines a extreme black
hole with mass

\begin{equation}
M_{max}=\sqrt[3]{\frac{3}{16}\frac{Q^{4}}{l^{2}}}=\sqrt[3]{-\frac{3}{16}Q^{4}\Lambda}\end{equation}
and with the horizons

\begin{eqnarray}
r_{1} & = & -2R=-2\sqrt{\frac{Ml^{2}}{3}}\\
r_{2} & = & r_{3}=R=\sqrt{\frac{Ml^{2}}{3}}.\end{eqnarray}
Note that $r_{1}$ is negative, so do not represent a physical horizon
(indeed it is always negative). Therefore, the black holes of case
I always have masses $M\leq M_{max}$ and only two physical horizons,
$r_{2}$ and $r_{3}$, that, for the extremal black hole, coincide.
The largest radius between $r_{2}$ and $r_{3}$ corresponds to the
event horizon of the black hole $r_{H}$, while the other corresponds
to the inner horizon.

\subsection{Heat Capacity }

The heat capacity of this black hole is defined by the relation

\begin{equation}
C_{Q}=\left(\frac{\partial M}{\partial T}\right)_{Q},\end{equation}
thus, using (\ref{eq:horizonmass}) we obtain

\begin{equation}
C_{Q}=4\pi r_{H}\left(\frac{12r_{H}^{3}-Q^{2}l^{2}}{12r_{H}^{3}+2Q^{2}l^{2}}\right).\end{equation}
Therefore, $C_{Q}$ is positive if

\begin{equation}
r_{H}^{3}-\frac{Q^{2}l^{2}}{12}>0\end{equation}

or, using again equation (\ref{eq:horizonmass}), we conclude that
the heat capacity is positive when the event horizon has a radius
$r_{H}$ that satisfies

\begin{equation}
r_{H}>\frac{Q^{2}}{4M}.\end{equation}

\section{The First Law of Thermodynamics}

In this section we will deduce the first law of thermodynamics for
the three-dimensional black hole with Coulomb-like electric field
using the field equations near the horizon. First, we will define
the thermodynamical quantities in terms of the mass of the black hole
given by equation (\ref{eq:horizonmass}). The surface gravity at
the horizon is

\begin{equation}
\kappa=\frac{1}{2}\left.\frac{df}{dr}\right|_{r=r_{H}}=\frac{r_{H}}{l^{2}}-\frac{Q^{2}}{12r_{H}^{2}}\end{equation}
and the Hawking temperature can be expressed as

\begin{equation}
T=\frac{\kappa}{2\pi}=\frac{r_{H}}{2\pi l^{2}}-\frac{Q^{2}}{24\pi r_{H}^{2}}.\end{equation}

The Bekenstein-Hawking entropy is given by

\begin{equation}
S=\frac{2\pi r_{H}}{4G},\end{equation}
where $2\pi r_{H}$ is the perimeter of the horizon. Since we are
working in units such that $G=\frac{1}{8}$, the entropy becomes twice
the perimeter,

\begin{equation}
S=4\pi r_{H}.\end{equation}
The electrostatic potential at the horizon is defined in terms of
the mass by

\begin{equation}
\Phi=\left.\frac{\partial M}{\partial Q}\right|_{r=r_{H}}=\frac{Q}{6r_{H}}.\end{equation}
Now, the Einstein tensor has non-zero components

\begin{eqnarray*}
G_{t}^{t}=G_{r}^{r} & = & \frac{1}{2r}\frac{df}{dr}\\
G_{\varphi}^{\varphi} & = & \frac{1}{r}\frac{d^{2}f}{dr^{2}}.\end{eqnarray*}

Therefore, the $\left(r,r\right)$ component of the field equations
is

\begin{equation}
G_{r}^{r}+\Lambda g_{r}^{r}=\pi T_{r}^{r}\end{equation}
and when this equation is evaluated at the horizon $r=r_{H}$, we
have

\begin{equation}
\left.\frac{df}{dr}\right|_{r=r_{H}}-\frac{2r_{H}}{l^{2}}=2\pi r_{H}T_{r}^{r},\end{equation}
where we have used $\Lambda=-\frac{1}{l^{2}}$. The component $T_{r}^{r}$
of the stress-energy tensor can be interpreted as the radial pressure
$\left(T_{r}^{r}=-P\right)$, then

\begin{equation}
\left.\frac{df}{dr}\right|_{r=r_{H}}-\frac{2r_{H}}{l^{2}}=-2\pi r_{H}P.\end{equation}
To give a thermodynamical interpretation of this equation, we consider
a virtual displacement of the horizon, $dr_{H}$. Thus, multiplying
on both sides of the equation by this factor, 

\begin{equation}
\left.\frac{df}{dr}\right|_{r=r_{H}}dr_{H}-\frac{2r_{H}}{l^{2}}dr_{H}=-2\pi r_{H}Pdr_{H}\end{equation}

\begin{equation}
\frac{1}{4\pi}\left.\frac{df}{dr}\right|_{r=r_{H}}d\left(4\pi r_{H}\right)-\frac{2r_{H}}{l^{2}}dr_{H}=-d\left(\pi r_{H}^{2}\right)P.\end{equation}

Using equation (\ref{eq:horizonmass}) as the definition of the mass
on the horizon, we obtain the differential

\begin{equation}
dM=\frac{2r_{H}}{l^{2}}dr_{H}+\frac{\Phi}{2}dQ,\end{equation}
and the field equation becomes

\begin{equation}
\frac{1}{4\pi}\left.\frac{df}{dr}\right|_{r=r_{H}}d\left(4\pi r_{H}\right)-dM+\frac{\Phi}{2}dQ=-d\left(\pi r_{H}^{2}\right)P.\end{equation}
The definition of entropy and Hawking temperature, gives

\begin{equation}
TdS-dM+\frac{\Phi}{2}dQ=-PdA,\end{equation}
where $A$ is the area enclosed by the horizon. Thus, the field equation
takes the form

\begin{equation}
dM+\frac{\Phi}{2}dQ=TdS+\Phi dQ+PdA.\end{equation}
 The extra term at the left hand side is just the electrostatic energy
enclosed by the horizon, and therefore, we identify 

\begin{equation}
M+\frac{\Phi}{2}Q=E\end{equation}
as the total energy inside the horizon, and the field equation takes
the usual form of the first law of thermodynamics,

\begin{equation}
dE=TdS+\Phi dQ+PdA.\end{equation}

\section{The $\Lambda=0$ Black Hole}

If we consider a zero cosmological constant, $\Lambda=0$, the resulting
black hole has interesting properties. The line element becomes 

\begin{equation}
ds^{2}=-\left(-M+\frac{Q^{2}}{6r}\right)dt^{2}+\frac{dr^{2}}{\left(-M+\frac{Q^{2}}{6r}\right)}+r^{2}d\varphi^{2},\end{equation}
that shows how this spacetime is asymptotically flat. This charged
black hole has just one horizon at

\begin{equation}
r_{H}=\frac{Q^{2}}{6M}.\end{equation}
The surface gravity at the horizon is given by

\begin{equation}
\kappa=-\frac{Q^{2}}{12r_{H}^{2}}=-\frac{3M^{2}}{Q^{2}}\end{equation}
and the Hawking temperature is negative,

\begin{equation}
T=-\frac{Q^{2}}{24\pi r_{H}^{2}}=-\frac{3M^{2}}{2\pi Q^{2}}.\end{equation}

The posibility of a negative temperature has been associated with
the existence of \emph{exotic matter}, but as has been shown by Cataldo
et. al. \cite{cataldo}, the non-linear electrodynamics used as source
in the field equation of this black hole satisfies the weak energy
condition. However, a possibility of negative temperatures without
exotic matter has been propposed in the de Sitter geometry \cite{balasu},
but due to the thermodynamical instability of this space, the negative
temperature is prohibited.  

However, one proposal that works fine is to consider, as recently
done by Arraut, et. al. \cite{arraut}, the surface gravity as the
absolute value

\begin{equation}
\kappa=\frac{1}{2}\left|\frac{df}{dr}\right|{}_{r=r_{H}}=\frac{Q^{2}}{12r_{H}^{2}}=\frac{3M^{2}}{Q^{2}}.\end{equation}
The reason for this election is the equivalence principle because,
in this case, the local inertial observer is moving while $r$ increases,
thus, a static observer has a negative scalar acceleration.

On the other hand, the heat capacity for the $\Lambda=0$ black hole
is given by 

\begin{equation}
C_{Q}=-\frac{\pi Q^{2}}{3M}=-2\pi r_{H},\end{equation}
i.e. minus the perimeter of the event horizon. Since the heat capacity
of this black hole is always negative, $\frac{\partial T}{\partial M}<0$,
the behavior of $T$ is the expected, i.e. as $M$ increases, the
temperature decreaces\@.

\section{Conclusion}

We have studied the thermodynamics of the $\left(2+1\right)$ dimensional
black hole with a Coulomb-like electric field, obtained by the use
of a non-linear electrodynamics. By considering a virtual displacement
of the horizon, we have shown that the field equations have a thermodynamical
interpretation since they can be rewritten as the differential form
of the first law, 

\begin{equation}
dE=TdS+\Phi dQ+PdA,\end{equation}
where $E$ is the total energy inside the horizon (that corresponds
to the mass of the black hole plus the electrostatic energy enclosed
by the horizon). This fact shows how the thermodynamical properties
are undoubtely related with the presence of horizons and maybe, they
are just a consequence of the holographic properties of gravity. A
further study on this area is in progress.

Finally, using a zero cosmological constant, $\Lambda=0$, the solution
becomes a black hole with a negative Hawking temperature and a negative
heat capacity even though the non-linear electric field used as source
satisfies the weak energy condition and does not behave as exotic
matter. In order to obtain a positive temperature, we use the equivalence
principle to consider the absolute value of the surface gravity as
the relevant quantity.

\end{document}